\UseRawInputEncoding
\documentclass[11pt,onecolumn]{amsart}
\usepackage[left=1.5cm,top=1.5cm,bottom=1.5cm,right=1.5cm]{geometry}
\usepackage{latexsym}
\usepackage{amssymb,amsmath,amsfonts}
\usepackage[pagewise]{lineno}
\usepackage{multirow}
\usepackage{graphicx}
\usepackage[colorlinks = true]{hyperref}
\theoremstyle{definition}

\newcommand{\N}{\mathbb N}

\numberwithin{equation}{section}
 \global\parskip 6pt \oddsidemargin
0.1cm \evensidemargin 0.4cm \headheight 0.1cm \topmargin -2.0cm
\begin{document}
\pagenumbering{arabic}
\title[Analyzing population dynamics models via Sumudu transform]{Analyzing population dynamics models via Sumudu transform}
\author{ M.O. Aibinu$^{1*}$, S. C. Thakur$^2$, S. Moyo$^3$}
\address{$^{1}$ Institute for Systems Science \& KZN e-Skills CoLab, Durban University of Technology, Durban 4000, South Africa}
\address{$^{1}$  DSI-NRF Centre of Excellence in Mathematical and Statistical Sciences (CoE-MaSS), Johannesburg, South Africa}
\address{$^{1}$ National Institute for Theoretical and Computational Sciences (NITheCS), South Africa}
\address{$^{2}$ KZN e-Skills CoLab, Durban University of  Technology, Durban 4000, South Africa}
\address{$^{3}$Institute for Systems Science \& Office of the DVC Research, Innovation \& Engagement, Milena Court, Durban University of Technology, Durban 4000, South Africa}
\email{$^*$moaibinu@yahoo.com / mathewa@dut.ac.za}


\begin{abstract}
This study demonstrates how to construct the solutions of a more general form of population dynamics models via a blend of variational iterative method with Sumudu transform. In this paper, population growth models are formulated in the form of delay differential equations of pantograph type which is a general form for the existing models. Innovative ways are presented for obtaining the solutions of population growth models where other analytic methods fail.  Stimulating procedures for finding patterns and regularities in seemingly chaotic processes have been elucidated in this paper.  How, when and why the changes in population sizes occur can be deduced through this study.
\end{abstract}

\maketitle
\section{Introduction}
\par There is a growing need to understand the dynamics that affect populations of organisms over time. The study of population dynamics can help to understand what influences the abundance of organisms at a particular time. It can also help to find answer to why the abundance of a species of organisms changes over time. Finding patterns and regularities in seemingly chaotic processes is through the studies on population dynamics. Prediction on certain species of animals or plants which are under the danger of extinction is as a result of studies on population dynamics. How a deadly virus spreads can only be understood and explained through population dynamics. There are several reports on using mathematical models to analyze the population dynamics \cite{Nickbakhsh1, Fischer1, Dong1, Brady1, Tong1, Ezz-Eldien}. Mathematical modeling of physical situations involves differential equations. Differential equations are indispensable for the description of many real-world phenomena. Differential equations are generally known as powerful tools for the study, analysis and prediction of essential real-world occurrences. 
\par Let $x(t)$ denote the population size at time $t$ and let $r$ be the population growth rate (birth rate minus death rate). In 1798, the studies of Malthus on population growth was summarized by the model \cite{Malthus}, 
 \begin{equation}\label{ppd1}
 x'(t)=rx(t), \ \ x(0)=x_0.
 \end{equation} 
Malthus was a British demographer and and came up with the model which is given by (\ref{ppd1}). The model was as a result of studies of demographic data which were more than a hundred years. Solving the initial value problem (\ref{ppd1}) by separating the variables yields
 \begin{equation}\label{ppd2}
 x(t)=x_0e^{rt}.
 \end{equation}
 Obviously, it is an exponential model. It speculates exponential growth if $r > 0$ and represents the traditional decay if $r < 0.$ Such a model may be valid for only a short period or only when the population is scarce and resources are abundant. It is a common knowledge that at a certain level of growth of a population, certain negative factors do set in which hinder further growth. Instances of such negative factors include poor agricultural yields which could cause starvation due to food shortages, air pollution as well as emergence of virus and diseases that could affect the lifespans of the organisms. Indeed, prediction of the model by Malthus is unrealistic in nature. In 1838, Verhulst considered the fact that resources are limited and proposed the logistic growth model \cite{Verhulst},
  \begin{equation}\label{ppd3}
x'(t)=rx(t)\left(1- \frac{x(t)}{K}\right),
 \end{equation}
where $r(> 0)$ is the intrinsic growth rate and $K(> 0)$ is the carrying capacity of the population. Given an initial condition $x_0,$ the solution for the logistic growth model is obtained as
 \begin{equation}\label{ppd4}
 x(t)=\frac{x_0K}{x_0+(K-x_0)e^{-rt}}.
 \end{equation}
 Competition occurs as $x(t)$ gets large and $x(t)$ approaches $K$ as $t\rightarrow \infty.$ The logistic growth model predicts a rapid growth when $x(t)$ is smaller than $K$ and it stipulates decrease of growth when $x(t)$ approaches $K.$ If $x_0 = K,$ the population remains in time at $x(t) = K,$ which is an equilibrium point. Naturally, occurrences of processes are not instantaneous. Behavioral responses of organisms to environmental changes takes a unit of time before it is feasible. Recovery of grasses after grazing takes a unit of time. Ordinary differential equations are not enough to model these kinds of scenarios. The growth rate $x'(t)$ at time $t$ depends on both $x(t)$ and $x(t-\tau),$  where $x(t-\tau)$ is the population size in some period in the past $(t - \tau )$ and $\tau >0$ is a constant. In 1948, Hutchinson proposed a more logistic growth model that involves delay \cite{Hutchinson},
 \begin{equation}\label{ppd5}
x'(t)=rx(t)\left(1- \frac{x(t-\tau)}{K}\right),  \ \ x(0)=x_0.
 \end{equation} 
However, the analytic solutions to  equations of the form (\ref{ppd5}) are not possible in general. Numerical approximations have been applied to gain an insight into the behavior of such models. Research efforts over the years have been on the construction of solutions, reformulations, improvements and applications of (\ref{ppd5}) \cite{Tong1, Ezz-Eldien, Brauer, Dawed, Galeano2}. Recently, the research focus has been on delay differential equations with multi-proportional delays. Generally, they are being referred to as delay differential equations of pantograph type \cite{Polyanin1, Hou, Jafari, Ali}. Delay differential equations of pantograph type have received much attention because many models with proportional delays produce significant results. The results which they produce are found to be accurate and stable. Also, they are suitable for studying several models which include the case where the delay is a fraction of the time unit. Consequently, several existing models and various classes of differential equations are being reformulated and reconstructed in terms of delay differential equations of pantograph type. 
\par The studies on population dynamics are of crucial importance as all of the processes on the earth, directly or indirectly affect the human life. The goal of this study is to present suitable reformulation and reconstruction for some existing population growth models in terms of delay differential equations of pantograph type. An innovative method is displayed for constructing the solutions of the models where other analytic methods fail. This study shows how to find patterns and regularities in seemingly chaotic processes. Some single and interacting species population models are illustrated graphically and analyzed.

\section{Description of the method}
The method is presented in this section to make this study a complete paper. The account of the method which is being presented has been discussed in \cite{Liu1, Vilu1, Aibinu6}.
\subsection{Variational iterative method} Due to its flexibility, consistency and effectiveness, variational iterative method is more preferred when compared to other well-known methods (See e.g, \cite{Wu1, Wu2} and references there in). The elevation which is a blend of variational iterative method with Laplace transform (See e.g, \cite{Wu1, Wu2}) and Sumudu transform (See e.g, \cite{Vilu1}) has also been considered. The Sumudu transform can be described as a mutation of Laplace transform. Sumudu transform has been proved to be a simple, effective and universal way for obtaining Lagrange multiplier. For a given property of Laplace transform, a corresponding property can be obtained for Sumudu transform and vice versa (See e.g, \cite{Watugala1, Belgacem1, Belgacem2}). Sumudu transform can be applied to a given function $f(t)$ which satisfies the Dirichlet conditions:
\begin{itemize}
	\item [(i)] it is single valued function which may have a finite number of isolated discontinues for $t>0.$
	\item [(ii)] it remains less than $be^{-a_0t}$ as $t$ approaches $\infty,$ where $b$ is a positive constant and $a_0$ is a real positive number.
\end{itemize}
Sumudu transform has been employed to obtain the solutions of differential equations which are of several forms \cite{Hussein, Golmankhaneh, Alomari, Nisar}. 
\subsection{Presentation of Sumudu transform}\label{sumud1}
The concept was proposed by Watugala \cite{Watugala1}.  It is an integral transform and it is suitable for solving several problems which are modeled by differential equations. Let $F(u)$ denote the Sumudu transform of a function $f(t).$ For all real numbers $t\geq 0,$ 
 \begin{equation}\label{sumud18}
F(u)=S\left[f(t)\right]=\int^{\infty}_{0}f(ut)e^{-t}dt.
 \end{equation}
The Sumudu transform for the integer order derivatives is expressed as 
 \begin{equation}\label{sumud19}
S\left[\frac{df(t)}{dt}\right]=\frac{1}{u}\left[F(u)-f(0)\right].
 \end{equation}
For the $n$-order derivative, the Sumudu transform is given as 
 \begin{equation}\label{sumud20}
S\left[\frac{d^nf(t)}{dt^n}\right]=\frac{1}{u^n}\left[F(u)-\displaystyle\sum_{k=0}^{n-1}u^k\frac{d^kf(t)}{dt^k}|_{x=0}\right].
 \end{equation}
Linearity of Sumudu transform can be easily established and Sumudu transform is also credited for preserving units and linear functions (See e.g, \cite{Watugala1, Belgacem1}). These essential properties make solving problems easy by using Sumudu transform without the need to resort to a new frequency domain. The focal points of the Sumudu transform are exhibited for a broad nonlinear problem
 \begin{equation}\label{sumud21}
\frac{d^nx(t)}{dt^n}+R\left[x(t)\right]+N\left[x(t)\right]=g(t),
 \end{equation}
 subject to the initial conditions
 \begin{equation}\label{sumud22}
x^{(k)}(0)=a_k,
 \end{equation}
 where $x^{(k)}(0)=\frac{d^kx(0)}{dt^k},$ $k=0, 1, ..., n-1,$ $R$ is a linear operator, $N$ is a nonlinear operator, $g(t)$ is a given continuous function and the highest order derivative is $\frac{d^nx(t)}{dt^n}.$
 \par Let $X(u)=S[x(t)],$ by taking the Sumudu transform of (\ref{sumud21}), its linear part is transformed into an algebraic equation of the form
 \begin{equation}\label{sumud23}
\frac{1}{u^n}X(u)-\displaystyle\sum_{k=0}^{n-1}\frac{1}{u^{n-k}}y^{(k)}(0)=S\left[g(t)-R\left[x\right]-N\left[x\right]\right].
 \end{equation}
 The sequence of iteration is deduced as
  \begin{equation}\label{sumud24}
X_{n+1}(u)=X_n(u)+{\varphi}(u)\left(\frac{1}{u^n}X_n(u)-\displaystyle\sum_{k=0}^{n-1}\frac{1}{u^{n-k}}x^{(k)}(0)-S\left[g(t)-R\left[x\right]-N\left[x\right]\right]\right),
 \end{equation}
 where ${\varphi}(u)$ is the Lagrange multiplier.  The classical variation operator is taken on both sides of (\ref{sumud24}) while $S\left[R\left[x\right]+N\left[x\right]\right]$ is considered as restricted term. This gives 
  \begin{equation}\label{sumud25}
\delta X_{n+1}(u)=\delta X_{n}(u)+{\varphi}(u)\frac{1}{u^n}\delta X_n(u),
 \end{equation}
 from which it is obtained that
 \begin{equation}\label{sumud26}
{\varphi}(u)=-u^n.
 \end{equation}
 Substitute (\ref{sumud26}) into (\ref{sumud24}) and take the inverse-Sumudu transform $S^{-1}$ of (\ref{sumud24}) to obtain
  \begin{eqnarray}\label{sumud27}
x_{n+1}(t)&=&x_n(t)+S^{-1}\left[-u^n\left(\frac{1}{u^n}X(u)-\displaystyle\sum_{k=0}^{n-1}\frac{1}{u^{n-k}}x^{(k)}(0)-S\left[g(t)-R\left[x\right]-N\left[x\right]\right]\right)\right],\nonumber\\
&=&x_1(t)+S^{-1}\left[u^n\left(S\left[g(t)-R\left[x(t)\right]-N\left[x(t)\right]\right]\right)\right],
 \end{eqnarray}
 where 
   \begin{eqnarray}\label{sumud28}
x_1(x)&=&S^{-1}\left[\displaystyle\sum_{k=0}^{n-1}u^{k}x^{(k)}(0)\right]\nonumber\\
&=&x(0)+x'(0)t+...+\frac{x^{n-1}(0)t^{n-1}}{(n-1)!}.
 \end{eqnarray}
\subsection{Variable coefficient nonlinear equation}
Suppose the broad nonlinear problem (\ref{sumud21}) contains variable coefficients which makes it to become
 \begin{equation}\label{sumud29}
 \frac{d^nx(t)}{dt^n}+\alpha R_1[x(t)] + \beta (t)R_2[x(t)]+N\left[x(t)\right]=g(t),
 \end{equation}
 where $\alpha$ is a constant and $\beta (t)$ is a variable coefficient, $R_1$ and $R_2$ are linear operators and other terms remain as defined in (\ref{sumud21}). The Sumudu transform of (\ref{sumud29}) is taken to get
   \begin{eqnarray}\label{sumud30}
X_{n+1}(u)&=&X_n(u)+{\varphi}(u)(\frac{1}{u^n}X_n(u)-\displaystyle\sum_{k=0}^{n-1}\frac{1}{u^{n-k}}x^{(k)}(0)\nonumber\\
&&-S\left[g(t)-\alpha R_1[x] - \beta (t)R_2[x]-N\left[x\right]\right]).
 \end{eqnarray}
Here, the restricted term is $S\left[\beta (t)R_2[x]+N\left[x\right]\right].$ Obtain the Lagrange multiplier ${\varphi}(u).$ The rest of computation processes are the same with the presentations in Section \ref{sumud1}. 


 \section{Main Results}
 \subsection{Modified Hutchinson's model} We consider the reconstruction of the model of Hutchinson (1948) in terms of delay differential equations of pantograph type to gain insight into situation which include the case where the delay is a fraction of the time unit. A suitable reformulation is presented to the Hutchinson's model (\ref{ppd5}). 
 \par Consider 
 \begin{eqnarray}\label{pd1}
x'(t)&=&rx(t)\left(1- \frac{x(\alpha t)}{K}\right),~t > 0,\\
 x(0)&=&x_0,\nonumber
 \end{eqnarray}
where $r$ and $K$ have the same meaning as in the previous equations and $\alpha \in [0, 1].$
\par Here, three case will be considered for the values of $\alpha$ which are $0, 1$ and $(0, 1).$
\par {\bf Case one:} $\alpha =0.$
\par
\begin{figure}
\includegraphics[width=12.0cm ,height=10.0cm]{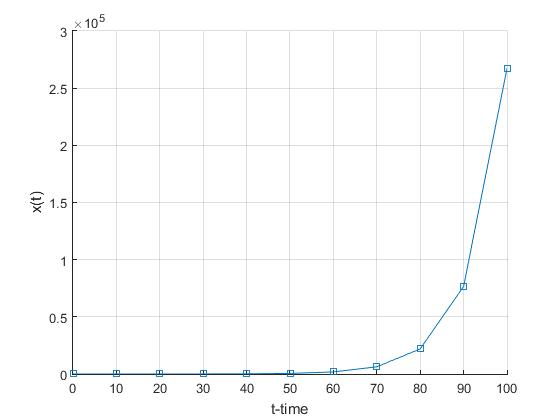}
\caption{Graph of $x(t)$ for $x_0=1, K=3000$ \& $r=0.125$.}
\label{ppd8}
\end{figure}

 For $\alpha =0,$ equation (\ref{pd1}) becomes
\begin{eqnarray}\label{ppd6}
x'(t)&=&rx(t)\left(1- \frac{x_0}{K}\right), \ \ x(0)=x_0.
 \end{eqnarray}
Equation (\ref{ppd6}) is a modified form that involves the carrying capacity of the population for the model of Malthus (1798).  The solution of (\ref{ppd6}) is given by
 \begin{equation}\label{ppd7}
 x(t)=x_0e^{r\left(1- \frac{x_0}{K}\right)t},
 \end{equation} 
 which is a modified {\bf exponential growth model}. Figure \ref{ppd8} displays the solution for modified Hutchinson's model when $\alpha =0.$
\par {\bf Case Two:} $\alpha =1.$
\par
\begin{figure}
\includegraphics[width=12.0cm ,height=10.0cm]{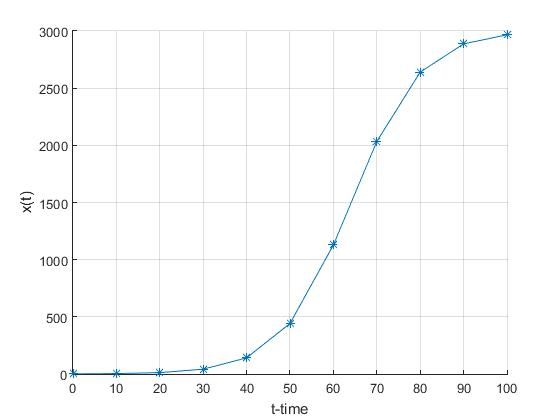}
\caption{Graph of $x(t)$ for $x_0=1, K=3000$ \& $r=0.125$.}
\label{ppd9}
\end{figure}

 Taking $\alpha =1$ in (\ref{pd1}) gives the logistic growth model (\ref{ppd3}) and its solution is given by (\ref{ppd4}). Figure \ref{ppd9} displays the solution for the logistic growth model where $x(t),$ which is the population size converges to $K,$ which is the carrying capacity of the population.
\par {\bf Case Three:} $\alpha \in (0, 1).$
\par The solution is constructed by using a blend of VIM with ST. Taking the ST of (\ref{pd1}) gives
\begin{equation}\label{pd6}
\frac{X(u)}{u}-\frac{x(0)}{u}=S\left[rx(t)\left(1- \frac{x(\alpha t)}{K}\right)\right].
\end{equation}
Since $x(0)=x_0,$ equation (\ref{pd6}) becomes
\begin{equation}
\frac{X(u)}{u}-\frac{x_0}{u}=S\left[rx(t)\left(1- \frac{x(\alpha t)}{K}\right)\right].
\end{equation}
 Thus for $ n\in \N,$ the variational iteration formula is given by
\begin{eqnarray}\label{pd2}
X_{n+1}(u)&=&X_n(u)+{\varphi}(u)\left(\frac{X_n(u)}{u}-\frac{x_0}{u}-S\left[rx_n(t)\left(1- \frac{x_n(\alpha t)}{K}\right)\right]\right).
\end{eqnarray}
The classical variation operator on both sides of (\ref{pd2}) is taken and the term $rx_n(t)\left(1- \frac{x_n(\alpha t)}{K}\right)$ is considered as the restricted variation. The Lagrange multiplier is then obtained as
\begin{equation}
{\varphi}(u)=-u.
\end{equation}
The inverse-Sumudu transform, $S^{-1}$ of (\ref{pd2}) is taken which gives the explicit iteration formula
\begin{eqnarray}\label{pd3}
x_{n+1}(t)&=&x_n(t)+S^{-1}\left[-u\left(\frac{X_n(u)}{u}-\frac{x_0}{u}-S\left[rx_n(t)\left(1- \frac{x_n(\alpha t)}{K}\right)\right]\right)\right]\nonumber\\
&=&x_0+S^{-1}\left[ru\left(S\left[x_n(t)\right]- \frac{1}{K}S\left[x_n(t)x_n(\alpha t)\right]\right)\right],
\end{eqnarray}
where the initial approximation is given by $x_1(t)=x(0)=x_0.$ Recall the decomposition of a nonlinear term $'N(x)'$ as $$N(x)=\displaystyle\sum_{i=0}^{\infty}A_i=\frac{1}{i!}\left[\frac{d^i}{d{\theta}^i}f\left(\displaystyle\sum_{n=0}^{\infty}{\theta}^nv_n\right)\right]\bigg|_{\theta=0},$$
 where $A_i$ is the Adomian polynomial \cite{Adomian}. Let $x_n=\displaystyle\sum_{i=0}^nv_i,$ the Adomian series of the nonlinear term $'x_n(t)x_n(\alpha t)'$ reads
\begin{equation}
\begin{cases}
A_0=v_0^2,\\
A_1=2v_0v_1,\\
A_2=2v_0v_2+v_1^2\\
A_3=2v_0v_3+2v_1v_2\\
\vdots
\end{cases}
\end{equation}
Therefore, this yields the successive formula
\begin{equation}\label{pd4}
\begin{cases}
v_0(t)=v(0)=x_0,\\
v_{n+1}(t)= S^{-1}\left[ru\left(S\left[v_n\right]- \frac{1}{K}S\left[A_n\right]\right)\right],
\end{cases}
\end{equation} 
which produces the iteration
\begin{equation}
\begin{cases}
v_0=x_0,\\
v_1=x_0\left(1- \frac{x_0}{K}\right)rt,\\
v_2=\left[x_0\left(1- \frac{x_0}{K}\right)^2-\frac{x_0^2}{K}\left(1- \frac{x_0}{K}\right)\right]\frac{r^2t^2}{2!},\\
\\
v_3=\bigg[ x_0\left(1- \frac{x_0}{K}\right)^3-\frac{4x_0^2}{K}\left(1- \frac{x_0}{K}\right)^2+\frac{x_0^3}{K}\left(1-\frac{x_0}{K}\right)\bigg]\frac{r^3t^3}{3!},\\
\vdots
\end{cases}
\end{equation}

\begin{figure}
\includegraphics[width=12.0cm ,height=10.0cm]{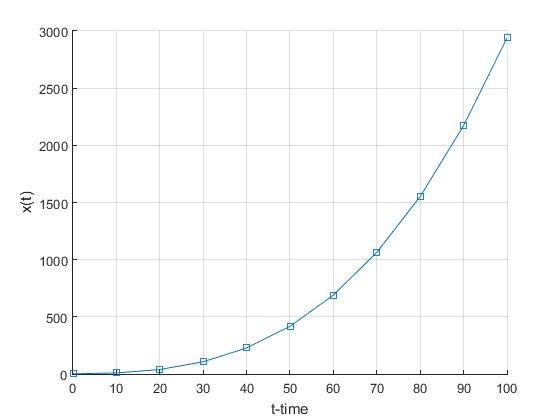}
\caption{Graph of $x(t)$ for $x_0=1, K=3000$ \& $r=0.125$.}
\label{pdd6}
\end{figure}

The solution is therefore given by
\begin{equation}\label{pd5}
\begin{split}
x(t) &=\displaystyle \lim_{n \rightarrow \infty} x_n= \displaystyle \lim_{n \rightarrow \infty} \displaystyle \sum_{i=0}^n v_i\\
&= x_0 + x_0\left(1- \frac{x_0}{K}\right)rt + \left[x_0\left(1- \frac{x_0}{K}\right)^2-\frac{x_0^2}{K}\left(1- \frac{x_0}{K}\right)\right]\frac{r^2t^2}{2!}\\
 &+\bigg[ x_0\left(1- \frac{x_0}{K}\right)^3-\frac{4x_0^2}{K}\left(1- \frac{x_0}{K}\right)^2+\frac{x_0^3}{K}\left(1-\frac{x_0}{K}\right)\bigg]\frac{r^3t^3}{3!}+ ...\\
&=x_0\left(1+\left(1- \frac{x_0}{K}\right)rt + \frac{r^2\left(1- \frac{x_0}{K}\right)^2t^2}{2!} + \frac{r^3\left(1- \frac{x_0}{K}\right)^3t^3}{3!} + ... \right)\\
&-\frac{x_0^2}{K}\left(1- \frac{x_0}{K}\right)\frac{r^2t^2}{2!}-\bigg[\frac{4x_0^2}{K}\left(1- \frac{x_0}{K}\right)^2-\frac{x_0^3}{K}\left(1-\frac{x_0}{K}\right)\bigg]\frac{r^3t^3}{3!}- ...\\
&=x_0e^{\left(1- \frac{x_0}{K}\right)rt}-\frac{x_0^2}{K}\left(1- \frac{x_0}{K}\right)\frac{r^2t^2}{2!}-\bigg[\frac{4x_0^2}{K}\left(1- \frac{x_0}{K}\right)^2-\frac{x_0^3}{K}\left(1-\frac{x_0}{K}\right)\bigg]\frac{r^3t^3}{3!}- ...
\end{split} 
\end{equation}
Figure \ref{pdd6} displays the solution for the modified Hutchinson's model when $\alpha \in (0, 1).$

\subsection{Blowflies model} The blowfly model was proposed in 1954 by Nicholson (See e.g, \cite{Arino1}). Gurney et al. (1980) \cite{Gurney1} introduced the delay into the model to correct the discrepancy which was noted in the Nicholson's blowflies model. The blowfly model with delay is given by
\begin{eqnarray}
x'(t)&=&Px(t-\tau)e^{\left(- \frac{x(t-\tau)}{x_0}\right)}-\delta x(t),~t > 0,\\
 x(0)&=&x_0,\nonumber
\end{eqnarray}
where $P$ is the maximum per capita daily egg production rate, $x_0$ is the size at which the blowflies population reproduces at its maximum rate and $\delta$ is the per capita daily adult death rate. To gain insight into situation  which
include the case where the delay is a fraction of the time unit, we study a reconstruction of blowfly model in the form of delay differential equations of pantograph type.
\par Consider a modified blowflies model which is expressed by 
 \begin{eqnarray}\label{nb1}
x'(t)&=&Px(\alpha t)e^{\left(- \frac{x(\alpha t)}{x_0}\right)}-\delta x(t),~t > 0,\\
 x(0)&=&x_0,\nonumber
 \end{eqnarray}
 where definitions of $P, x_0$ and $\delta$ remain the same while $\alpha \in [0, 1].$
 \par {\bf Case One:} $\alpha=0.$
 \par
\begin{figure}
\includegraphics[width=12.0cm ,height=10.0cm]{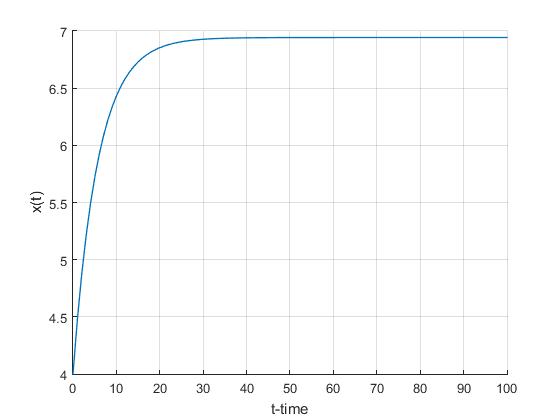}
 \caption{Graph of $x(t)$ for $x_0=4, P=2$ \& $\delta=0.175$.}
 \label{nbc2}
\end{figure}

 For $\alpha=0,$ equation (\ref{nb1})
$$x'(t)+\delta x(t)=Px_0e^{-1},~ x(0)=x_0,$$
which is a linear ordinary differential equation of degree one. Its solution is given 
\begin{equation}\label{nbc1}
x(t)=x_0\left(Pe^{-1}+\frac{1-Pe^{-1}}{e^{\delta t}}\right).
\end{equation}
An obvious deduction from (\ref{nbc1}) is that $x(t)\rightarrow Px_0e^{-1}$ as $t\rightarrow \infty.$ Figure \ref{nbc2} displays the solution for the modified blowflies model in the form of delay differential equations of pantograph type when $\alpha=0.$
\par {\bf Case Two:} $\alpha \in (0, 1].$
\par When $\alpha =1,$ observe that it is not easy to obtain by analytical method, the solution of the nonlinear first order differential equation
$$x'(t)-\left(Pe^{\left(- \frac{x(t)}{x_0}\right)}-\delta\right) x(t)=0,~ x(0)=x_0.$$
Therefore, a blend of VIM with ST will be used to obtain the solution when $\alpha \in (0, 1].$
\par {\bf Solution:} By taking the ST of (\ref{nb1}), we obtain
\begin{equation}\label{nb2}
\frac{X(u)}{u}-\frac{x(0)}{u}=PS\left[x(\alpha t)e^{\left(- \frac{x(\alpha t)}{x_0}\right)}\right]-\delta S\left[x(t)\right].
\end{equation}
Since $x(0)=x_0,$ equation (\ref{nb2}) gives
\begin{equation}\label{nb3}
\frac{X(u)}{u}-\frac{x_0}{u}=PS\left[x(\alpha t)e^{\left(- \frac{x(\alpha t)}{x_0}\right)}\right]-\delta S\left[x(t)\right].
\end{equation}
 Thus for $ n\in \N,$ the variational iteration formula is given by
\begin{eqnarray}\label{nb4}
X_{n+1}(u)&=&X_n(u)+{\varphi}(u)\left(\frac{X_n(u)}{u}-\frac{x_0}{u}-PS\left[x_n(\alpha t)e^{\left(- \frac{x_n(\alpha t)}{x_0}\right)}\right]+\delta S\left[x_n(t)\right]\right).
\end{eqnarray}
The classical variation operator on both sides of (\ref{nb4}) is taken and the term $Px(\alpha t)e^{\left(- \frac{x(\alpha t)}{x_0}\right)}$ is considered as the restricted variation. The Lagrange multiplier is then obtained as
\begin{equation}
{\varphi}(u)=-u.
\end{equation}
Taking the inverse-Sumudu transform, $S^{-1}$ of (\ref{nb4}) gives the explicit iteration formula
\begin{eqnarray}\label{nb5}
x_{n+1}(t)&=&x_n(t)+S^{-1}\left[-u\left(\frac{X_n(u)}{u}-\frac{x_0}{u}-PS\left[x_n(\alpha t)e^{\left(- \frac{x_n(\alpha t)}{x_0}\right)}\right]+\delta S\left[x_n(t)\right]\right)\right]\nonumber\\
&=&x_0+S^{-1}\left[u\left(PS\left[x_n(\alpha t)e^{\left(- \frac{x_n(\alpha t)}{x_0}\right)}\right]-\delta S\left[x(t)\right]\right)\right],
\end{eqnarray}
where the initial approximation is given by $x_1(t)=x(0)=x_0.$ Recall the decomposition $$N(x)=\displaystyle\sum_{i=0}^{\infty}A_i=\frac{1}{i!}\left[\frac{d^i}{d{\theta}^i}f\left(\displaystyle\sum_{n=0}^{\infty}{\theta}^nv_n\right)\right]\bigg|_{\theta=0},$$
 where $N(x)$ is the nonlinear term and $A_i$ is the Adomian polynomial \cite{Adomian}. Let $x_n=\displaystyle\sum_{i=0}^nv_i,$ and observe that the Adomian series of the nonlinear term $x(\alpha t)e^{\left(- \frac{x(\alpha t)}{x_0}\right)}$ reads
\begin{equation}\label{nb6}
\begin{cases}
A_0=v_0e^{-1},\\
A_1=v_1e^{-1},\\
A_2=v_2e^{-1},\\
A_3=v_3e^{-1},\\
\vdots
\end{cases}
\end{equation}
Therefore, this yields the successive formula
\begin{equation}\label{nb7}
\begin{cases}
v_0(t)=v(0)=x_0,\\
v_{n+1}(t)= S^{-1}\left[u\left(PS\left[A_n\right]-\delta S\left[v_n(t)\right]\right)\right],
\end{cases}
\end{equation} 
which produces the iteration
 \begin{equation}\label{nb8}
\begin{cases}
v_0=x_0,\\
v_1=x_0\left(Pe^{-1} - \delta \right)t,\\
v_2=x_0\frac{\left(Pe^{-1} - \delta \right)^2t^2}{2!},\\
v_3=x_0\frac{\left(Pe^{-1} - \delta \right)^3t^3}{3!},\\
v_4=x_0\frac{\left(Pe^{-1} - \delta \right)^4t^4}{4!},\\
\vdots
\end{cases}
\end{equation}

\begin{figure}
\includegraphics[width=12.0cm ,height=10.0cm]{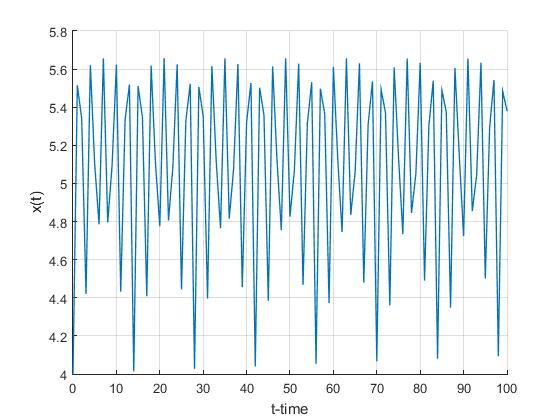}
 \caption{Graph of $x(t)$ for $x_0=4, P=2$ \& $\delta=0.175$.}
 \label{nbc3}
\end{figure}

The solution is therefore given by
\begin{equation}\label{nb9}
\begin{split}
x(t) &=\displaystyle \lim_{n \rightarrow \infty} x_n= \displaystyle \lim_{n \rightarrow \infty} \displaystyle \sum_{i=0}^n v_i\\
&= x_0\left(1+ \left(Pe^{-1} - \delta \right)t + \frac{\left(Pe^{-1} - \delta \right)^2t^2}{2!} + \frac{\left(Pe^{-1} - \delta \right)^3t^3}{3!} + ...\right)\\
&=x_0e^{\left(Pe^{-1} - \delta \right)t}\\
&=x_0e^{-\left(\delta-Pe^{-1}\right)t}\\
&=x_0\left(cos \left(\delta-Pe^{-1}\right)t - sin \left(\delta-Pe^{-1}\right)t \right).
\end{split} 
\end{equation}
Figure \ref{nbc3} displays the solution for the modified blowflies model in the form of delay differential equations of pantograph type when $\alpha \in (0, 1].$ The solution is in agreement with the results of Gurney et al. \cite{Gurney1}.
 
\vspace{1.0cm}
\noindent{\bf Conclusion}: This study presents the evolution of population growth models and reasons for the need of models which are in the form of delay differential equations of pantograph type. Suitable transformation for the existing models have been proposed in this paper. It is generally difficult to obtain the analytic solutions of the population growth models which involve the delays. This paper shows innovative ways for obtaining the solution where other analytic methods fail. The solutions are presented in this paper to both the existing and modified models. The modified models which have been presented in this paper are generalized forms for some existing models which were obtained by taking cases. Stimulating procedures for  finding patterns and regularities in seemingly chaotic processes have been elucidated. Population growth models for some of single and interacting species have been analyzed and illustrated by graphs. How, when, and why the changes in population sizes occur can be deduced from this study. This study provides information on effective ways for evaluating the impact of the physical environment on the species of organism and it can help to make accurate prediction on the population growth. This study can assist the conservation practitioners to evaluate the impact of the physical environment on a species and to determine whether the population in a given area will increase or decrease.

\vspace{1.0cm}
\noindent {\bf Acknowledgments}: The first author acknowledges with thanks the postdoctoral fellowship and financial support from the DSI-NRF Center of Excellence in Mathematical and Statistical Sciences (CoE-MaSS). Opinions expressed and conclusions arrived are those of the authors and are not necessarily to be attributed to the CoE-MaSS.

\end{document}